%% file: robust_vae.tex
\documentclass{article}
\usepackage{spconf,amsmath,graphicx}
\usepackage{booktabs}

\usepackage{amsmath,epsfig,amssymb,amsthm,url,amsfonts}
\usepackage{algorithm}
\usepackage{algpseudocode}

\usepackage{multirow}
\usepackage{mdframed}
\usepackage{url}
\usepackage{enumitem}
\usepackage[makeroom]{cancel}
\usepackage{dblfloatfix}
\usepackage{array}
\usepackage{epstopdf}
\usepackage{float}
\usepackage{subfig}
\usepackage{breqn}
\usepackage{cite}
\usepackage{xcolor}
\usepackage{tabularx}
\usepackage[printonlyused,withpage]{acronym}
\usepackage{balance}

\usepackage{tikz}
\usetikzlibrary{bayesnet,calc,tikzmark}
\usetikzlibrary{shapes.geometric}
\tikzset{  net/.style={draw,trapezium,trapezium angle=75,shape border rotate=270} }
\tikzset{  rnn/.style={draw,rectangle} }

\theoremstyle{break}

\theoremstyle{definition}

\newcommand{\compresslist}{
	\setlength{\itemsep}{1pt}
	\setlength{\parskip}{0pt}
	\setlength{\parsep}{0pt}
}

\newcommand{\bs}{\boldsymbol}

\newcommand{\vect}[1]{\mathbf{#1}}

\def\bs#1{\boldsymbol{#1}}

\title{Switching Variational Auto-Encoders for\\ Noise-Agnostic Audio-visual Speech Enhancement\vspace{-3mm}}
\name{Mostafa Sadeghi$^1$ and Xavier Alameda-Pineda,$^2$ IEEE Senior Member\thanks{Xavier Alameda-Pineda acknowledges ANR JCJC ML3RI project (ANR-19-CE33-0008-01). This work has been partially supported by MIAI @ University Grenoble Alpes, (ANR-19-P3IA-0003)}\vspace{-3mm}}
\address{$^1$Inria Nancy Grand-Est, $^2$Inria Grenoble Rh\^{o}ne-Alpes \& Univ. Grenoble Alpes, France\vspace{-3mm}}

\graphicspath{{../figs/}}

 \input{Header}

\acrodef{STOI}{short-time objective intelligibility}
\acrodef{SE}{speech enhancement}
\acrodef{STFT}{short-time Fourier transform}
\acrodef{PSD}{power spectral density}
\acrodef{NMF}{nonnegative matrix factorization}
\acrodef{AV}{audio-visual}
\acrodef{DNN}{deep neural network}
\acrodefplural{DNNs}{deep neural networks}
\acrodef{VAE}{variational auto-encoder}
\acrodefplural{VAEs}{variational auto-encoders}
\acrodef{CVAE}{conditional variational auto-encoder}
\acrodefplural{CVAEs}{conditional variational auto-encoders}
\acrodef{A-VAE}{audio VAE}
\acrodef{V-VAE}{visual VAE}
\acrodef{AV-CVAE}{audio-visual CVAE}
\acrodef{ROI}{region of interest}
\acrodef{MCMC}{Markov Chain Monte Carlo}
\acrodef{EM}{expectation-maximization}
\acrodef{VEM}{variational expectation-maximization}
\acrodef{MCEM}{Monte Carlo expectation-maximization}
\acrodef{TF}{time frequency}
\acrodef{ELBO}{evidence lower bound}
\acrodef{ROI}{region of interest}
\acrodef{LR}{Living Room}
\acrodef{SDR}{signal-to-distortion ratio}
\acrodef{PESQ}{perceptual evaluation of speech quality}
\acrodef{ASE}{audio speech enhancement}
\acrodef{VSE}{visual speech enhancement}
\acrodef{AVSE}{audio-visual speech enhancement}
\acrodef{SNR}{signal-to-noise ratio}
\acrodefplural{SNRs}{signal-to-noise ratios}
\acrodef{LSTM}{long short-term memory}
\acrodef{HMM}{hidden Markov model}
\acrodef{SwVAE}{switching variational auto-encoder}

\begin{document}
%
\maketitle
\begin{abstract}
Recently, audio-visual speech enhancement has been tackled in the unsupervised settings based on \acp{VAE}, where during training only clean data is used to train a generative model for speech, which at test time is combined with a noise model, e.g.\ \ac{NMF}, whose parameters are learned without supervision. Consequently, the proposed model is agnostic to the noise type. When visual data are clean, audio-visual VAE-based architectures usually outperform the audio-only counterpart. The opposite happens when the visual data are corrupted by clutter, e.g.\ the speaker not facing the camera. In this paper, we propose to find the optimal combination of these two architectures through time. More precisely, we introduce the use of a latent sequential variable with Markovian dependencies to switch between different VAE architectures through time in an unsupervised manner: leading to \ac{SwVAE}. We propose a variational factorization to approximate the computationally intractable posterior distribution. We also derive the corresponding variational expectation-maximization algorithm to estimate the parameters of the model and enhance the speech signal. Our experiments demonstrate the promising performance of \ac{SwVAE}.


\end{abstract}
\section{Introduction}
Audio-visual speech enhancement (AVSE) refers to the task of removing background noise from a noisy speech with the help of visual information (lip movements) of the unknown speech~\cite{girin2001audio,MichTZXYYJ20}. Several \ac{DNN}-based methods have been proposed for AVSE in the past. The majority of these methods are \textit{supervised}, where the underlying idea is to learn a \ac{DNN} that maps noisy speech and its associated visual data (video frames of mouth area) to clean speech \cite{hou2018audio,AfouCZ18,MichTZXYYJ20,GabbSP18}. To have a good generalization performance, a huge dataset with different noise types and various \ac{SNR} levels is usually required. 

Recently, some \textit{unsupervised} {AVSE} methods have been proposed that do not need noise signals for training \cite{sadeghiLAGH19,SadeA19a,SadeA20MinVAE}, meaning that their training is agnostic to the noise type. This approach builds upon the audio-only speech enhancement counterpart \cite{Leglaive_MLSP18,bando2018statistical} consisting of two main steps. First, modeling the probabilistic generative process of clean speech using \acp{VAE}~\cite{KingW14}. Second, combining it with a noise model, e.g.\ \ac{NMF}, to perform speech enhancement from noisy speech. 

One critical issue with {AVSE} methods, shared with other AV-processing tasks such as speaker localisation and tracking~\cite{cech2013active,ban2017exploiting}, is how to robustly handle noisy visual data at test time, e.g., when mouth area is heavily occluded or non-frontal. Exploiting such noisy visual data by an AVSE model trained on clean data may degrade the performance. In the supervised settings, this problem is usually addressed by proper data augmentation and efficient audio-visual fusion strategies during model training. For example, \cite{TriaCZ19} proposes to combine speaker embedding with visual cues to achieve more robustness to occluded visual stream. Moreover, during training, some artificial occlusions are added to video frames. In the VAE-based unsupervised settings, a totally different perspective is pursued owning to its probabilistic nature. In this regard, a robust generative model has been proposed in \cite{SadeA19a} which is a mixture of trained audio-based (A-VAE) and audio-visual based (AV-VAE) model. As such, following a variational inference approach, for noisy visual data the A-VAE model is chosen, whereas for clean visual data the AV-VAE model is used, thus providing robustness.

In this paper, we build upon \cite{SadeA19a} and introduce a new model and associated robust AVSE algorithm, where a Markovian dependency is assumed to switch between different VAE-based generative models, and term them \acf{SwVAE}. Alternatively, the proposed model can be understood as a \ac{HMM} \cite{bishop06} with emission probabilities given by the decoder of  several \acp{VAE}. Furthermore, we propose a variational factorization of the posterior distribution of the latent variables, enabling efficient inference and algorithm initialization. Experimental results demonstrate the superior performance of the proposed method compared to \cite{SadeA19a}.

The rest of the paper is organized as follows. Section~\ref{sec:prop} introduces the proposed \ac{SwVAE}. The inference and speech enhancement methodologies, and the relation of the present work to \cite{SadeA19a} are also detailed in this section. Section~\ref{sec:exp} presents and discusses the experiments.
\section{Switching Variational Autoencoders}\label{sec:prop}
In this section, we present a generative model for \ac{STFT} time frames of clean speech consisting of audio-only and audio-visual VAE models plus a switching variable deciding which model to be used for each audio frame. The switching variable is modeled with an \ac{HMM}. We also discuss how to structure the variance of the background noise via \ac{NMF}. Then, a variational approximation is proposed to estimate the model parameters and infer the latent variables, including the clean speech signal, from the noisy mixture.
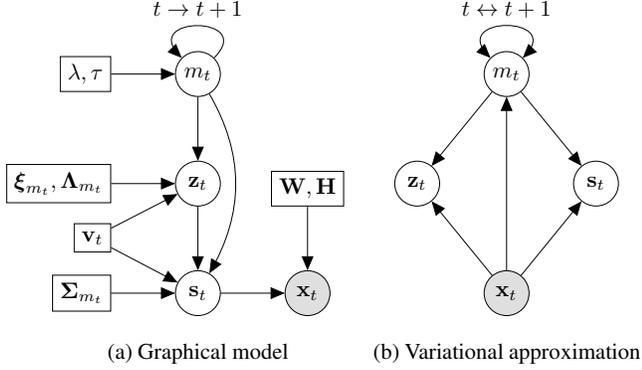
\begin{figure}[t]
 \resizebox{\columnwidth}{!}{
\begin{tikzpicture}
\node[latent] (m) {$m_t$};
\node[above=of m,yshift=-6mm] {$t\rightarrow t+1$};
\node[latent,below=of m] (z) {$\vect{z}_t$};
\node[latent,below=of z] (s) {$\vect{s}_t$};
\node[below=of s,yshift=7mm] {\text{(a) Graphical model}};
\node[obs,right=of s] (x) {$\vect{x}_t$};
\node[rnn,left=of m] (parm) {$\lambda,\tau$};
\node[rnn,left=of z] (parz) {$\boldsymbol{\xi}_{m_t},\boldsymbol{\Lambda}_{m_t}$};
\node[rnn,left=of s] (pars) {$\boldsymbol{\Sigma}_{m_t}$};
\node[rnn,left=of z,yshift=-8.5mm] (v) {$\vect{v}_{t}$};
\node[rnn,above=of x] (parx) {$\vect{W},\vect{H}$};
\edge{m}{z};
\edge{z}{s};
\edge{s}{x};
\edge{v}{z,s};
\edge{parz}{z};
\edge{pars}{s};
\edge{parm}{m};
\edge{parx}{x};
\path(m) edge[->,bend left=30] (s);
\draw[->] (m) to [out=45,in=135,loop,looseness=4.8] (m);
\node[obs,right=of x,xshift=14mm] (xinf) {$\vect{x}_t$};
\node[latent,above=of xinf,xshift=-14mm] (zinf) {$\vect{z}_t$};
\node[latent,above=of xinf,xshift=14mm] (sinf) {$\vect{s}_t$};
\node[latent,above=of zinf,xshift=14mm] (minf) {$m_t$};
\node[above=of minf,yshift=-6mm] {$t\leftrightarrow t+1$};
\node[below=of xinf,yshift=7mm] {\text{(b) Variational approximation}};
\edge{xinf}{minf,zinf,sinf};
\edge{minf}{zinf,sinf};
\draw[<->] (minf) to [out=45,in=135,loop,looseness=4.8] (minf);
\end{tikzpicture}}
\caption{Graphical model (left) and proposed variational inference (right) assotiacted to switching variational autoencoders.}\label{fig:model}
\end{figure}
\subsection{The generative model of SwVAE}
We define $ \vect{s}_t\in \mathbb{C}^F $ as the vector of clean speech \ac{STFT} coefficients at time frame $t \in \{1,...,T\}$. In the following, $\mathcal{N}_c$ and $\mathcal{N}$ stand for complex- and real-valued Gaussian distributions, respectively. The main methodological contribution of this paper is the use of a switching variable $m_t\in\{1,\ldots,M\}$ modeled with a Markov chain in combination with a set of $M$ non-linear generative models (i.e.\ VAE) to model clean speech. The full generative model describes the probabilistic relationship between the switching variable $m_t$, the clean speech $\vect{s}_t$, and the latent code $\vect{z}_t\in\mathbb{R}^L$, describing some hidden characteristics of $\vect{s}_t$, given the associated visual data representation $\vect{v}_t\in\mathbb{R}^V$. There are two possible, equivalent interpretations of this model. First, a hidden Markov model with emission probabilities given by the decoder of $M$ VAEs. Second, a set of $M$ VAEs switched by a selecting variable modeled with Markovian dependencies. More formally:
\begin{equation}
\begin{cases}
p(m_1,\ldots,m_T) \sim {\cal MC}(\lambda,\tau),\\p(\vect{z}_t|m_t;\vect{v}_t)\sim {\cal N}\Big(\boldsymbol{\xi}_{m_t}(\vect{v}_t),\vect{\Lambda}_{m_t}(\vect{v}_t)\Big),\\
p(\vect{s}_t|\vect{z}_t,m_t;\vect{v}_t)\sim {\cal N}_c\Big(\boldsymbol{0},\vect{\Sigma}_{m_t}(\vect{z}_t,\vect{v}_t)\Big),
\end{cases}
\label{eq:gen-model}
\end{equation}
where ${\cal MC}(\lambda,\tau)$ is short for a Markov chain with initial distribution $\lambda$ and transition distribution $\tau$, and $\boldsymbol{\xi}_{m_t}(.)$, $\vect{\Lambda}_{m_t}(.)$, and $\vect{\Sigma}_{m_t}(.,.)$ are non-linear transformations of their inputs indexed by $m_t\in\{1,\ldots,M\}$ and realized as \ac{DNN}s. For each generative model, the associated DNNs are trained by approximating the intractable posterior $p(\vect{z}_t|\vect{s}_t,m_t;\vect{v}_t)$ by another DNN-based parameterized Gaussian distribution called the encoder \cite{KingW14,sadeghiLAGH19}. So, there are $M$ different distributions for the prior of $ \vect{z}_t $ and for the likelihood of $ \vect{s}_t $. Importantly, the switching variable $m_t$ selects which one of the $M$ models is used at each time step $t$, while ensuring temporal smoothing in the choice of this transformation. To complete the definition of the probabilistic model, we use an \ac{NMF} structure for the additive noise \cite{bando2018statistical,Leglaive_MLSP18,sadeghiLAGH19}:
\begin{equation}
p(\vect{x}_t|\vect{s}_t) \sim \mathcal{N}_c\Big(\vect{s}_t, \text{diag}\Big(\Wb\bs{h}_t\Big)\Big),
\label{eq:nmf}
\end{equation}
where $ \Wb\in\Rbb_+^{F\times K}, \Hb\in\Rbb_+^{K\times T} $, and $ \bs{h}_t $ denotes the $ t $-th column of $ \Hb $. The graphical representation of the full model is shown in Fig.~\ref{fig:model}~(a). The set of HMM and NMF parameters, i.e. $ \lk \lambda, \tau, \Wb, \Hb \rk $ are then estimated following a variational inference method detailed in the next section, and represented in Fig.~\ref{fig:model}~(b). While for the generative model the dependencies are forward in time, at inference time, the latent code and spectrogram at any time $t$ depend on the past and future noisy observations. It should be emphasized that the DNN parameters of \eqref{eq:gen-model}, trained according to \cite{sadeghiLAGH19}, are fixed.


\subsection{Variational Inference}
In the proposed formulation, the problem of speech enhancement is cast into the computation of the posterior probability $p(\vect{s}|\vect{x},\vect{v})$, which is the marginal of the full posterior $p(\vect{s},\vect{z},\vect{m}|\vect{x},\vect{v})$, where we define $ \vect{x} =\lk \vect{x}_t \rk_{t=1}^T $ and analogously $ \vect{s},\vect{z},\vect{m},\vect{v} $. The full posterior being intractable, we propose the following variational factorization:
\begin{equation}
p(\vect{s},\vect{z},\vect{m}|\vect{x},\vect{v}) \approx r^s(\vect{s}|\vect{m})r^z(\vect{z}|\vect{m})r^m(\vect{m}).
\label{eq:var-dists}
\end{equation}
It is easy to see that $r^s$ and $r^z$ further factorize over time, meaning that: $r^s(\vect{s}|\vect{m}) = \prod_t r^s(\vect{s}_t|m_t)$ and analogously for $r^z(\vect{z}|\vect{m})$. Moreover, as a variational approximation, the posterior of the latent code $\vect{z}_t$ is assumed to follow a Gaussian distribution $r^z(\vect{z}_t|{m}_t) = {\cal N}(\vect{c}_{tm},\vect{\Omega}_{tm})$, where the mean vector $ \vect{c}_{tm} $ and the diagonal covariance matrix $ \vect{\Omega}_{tm} $ are to be estimated along with $ r^s $ and $ r^m $. To this end, we optimize the following lower-bound of the data log-likelihood $ \log p(\vect{x},\vect{v}) $, as done in variational inference:
\begin{equation}
\mathbb{E}_{r^s r^z r^m}\left[\log \frac{p(\vect{x},\vect{v},\vect{s},\vect{z},\vect{m})}{r^s(\vect{s}|\vect{m})r^z(\vect{z}|\vect{m})r^m(\vect{m})}  \right] \le \log p(\vect{x},\vect{v}).
\label{eq:elbo}
\end{equation}
\subsubsection{E-s step}
Optimizing \eqref{eq:elbo} over $r^s$ provides the following expression:
\begin{equation*}
r^s(\vect{s}_t|{m}_t)\propto p(\vect{x}_t|\vect{s}_t)\cdot\exp\Big(\mathbb{E}_{r^z}\Big[ \log p(\vect{s}_t|\vect{z}_t,{m}_t;\vect{v}_t) \Big]\Big).
\end{equation*}
Approximating the intractable expectation with a Monte-Carlo estimate, we obtain a Gaussian distribution: $r^s(\vect{s}_t|{m}_t)=\mathcal{N}_c(\bs{\eta}^{m_t}_t,\mbox{diag}[\bs{\nu}^{m_t}_t])$, where:
\begin{equation}
\label{eq:rs-params}
\eta_{ft}^{m_t} = \frac{\gamma_{ft}^{m_t}}{\gamma_{ft}^{m_t} + \left(\mathbf{W}\mathbf{H}\right)_{ft}}\cdot x_{ft},~
\nu_{ft}^{m_t} = \frac{\gamma_{ft}^{m_t}\cdot \left(\mathbf{W}\mathbf{H}\right)_{ft}}{\gamma_{ft}^{m_t} + \left(\mathbf{W}\mathbf{H}\right)_{ft}},
\end{equation}
\begin{equation}
\label{eq:gamma_}
\gamma_{ft}^{m_t} =\Big[\frac{1}{D}\sum_{d=1}^{D} {\Sigma_{m_t,ff}^{-1}(\zb_{{m}_t}^{(d)},\vb_t)}\Big]^{-1},
\end{equation}
in which, $\Sigma_{m_t,ff}$ denotes the $(f,f)$-th entry of $\vect{\Sigma}_{m_t}$ (similarly for the rest of the variables), and $\{\zb_{{m}_t}^{(d)}\}_{d=1}^D$ is a sequence sampled from $ r^z(\zb_t|{m}_t) $. The result in~(\ref{eq:rs-params}) must be interpreted as a Wiener filter, averaged over the latent variable $\vect{z}_t$ for a given VAE generative model $m_t$. The enhanced speech signal is the marginalisation over the switching variable at time $t$, and naturally writes:
\begin{equation}
\hat{\vect{s}}_t = \mathbb{E}_{r^m(m_t)}\Big[ \mathbb{E}_{r^s(\vect{s}_t|{m}_t)}[ \vect{s}_t  ]\Big]=\sum_{m_t} r^m(m_t) \bs{\eta}^{m_t}_t,~~~\forall t.
\label{eq:se}
\end{equation}

\subsubsection{E-z step}
After doing some derivations, the set of parameters of $r^z(\vect{z}_t|{m}_t)$ is estimated by solving:
\begin{align}
\label{eq:rz-update}
\max_{\vect{c}_{tm},\vect{\Omega}_{tm}}& \mathbb{E}_{r^m({m}_t)}\Big[ \mathbb{E}_{r^z(\vect{z}_t|{m}_t)}\Big[ \mathbb{E}_{r^s(\vect{s}_t|{m}_t)}\Big[ \log p(\vect{s}_t|\vect{z}_t,{m}_t;\vect{v}_t)\Big] \Big]\nonumber\\ &-\textrm{KL}(r^z(\vect{z}_t|{m}_t)\|p(\vect{z}_t|{m}_t;\vect{v}_t))\Big].
\end{align}
where, $ \textrm{KL} $ denotes the Kullback-Leibler
divergence. In \eqref{eq:rz-update}, the expectation over $ r^m $ and $ r^s $ can be evaluated in closed-form. This is also the case for the KL term as both the distributions are Gaussian. However, the expectation over $ r^z $ is intractable. Like in standard VAE, here we approximate this expectation with a single sample drawn from $ r^z $. Furthermore, to be able to back-propagate through the posterior parameters, the reparametrization trick is utilized \cite{KingW14}.
\subsubsection{E-m step}
For $r^m(\vect{m})$, we obtain:
\begin{equation}
r^m(\vect{m}) \propto p(\vect{m})\cdot \prod_{t=1}^T\exp(-g_t(m_t))
\label{eq:rm-update}
\end{equation}
with:
\begin{align}
g_t({m}_t)=&\mathbb{E}_{r^z}\Big[\textrm{KL}(r^s(\vect{s}_t|{m}_t)\|p(\vect{s}_t|\vect{z}_t,{m}_t;\vect{v}_t))\Big]-\label{eq:gm}\\&\mathbb{E}_{r^s}\Big[\log p(\vect{x}_t|\vect{s}_t) \Big]+\textrm{KL}(r^z(\vect{z}_t|{m}_t)\|p(\vect{z}_t|{m}_t;\vect{v}_t))\nonumber
\end{align}
Again, the KL terms and the expectation over $ r^s $ can be computed in closed-form. However, we approximate the expectation over $ r^z $ by a Monte-Carlo estimate. This allows us to compute~(\ref{eq:gm}). In order to compute the marginal variational posterior $r^m(m_t)$ required in the E-s and E-z steps, we realize that~(\ref{eq:rm-update}) has the same structure as standard HMM if we consider $\exp(-g_t(m_t))$ as the emission probability of the HMM. We therefore use the forward-backward algorithm \cite{bishop06} to efficiently compute $r^m(m_t)$.

\subsubsection{M step}
After performing the E steps, the NMF parameters are updated by optimizing \eqref{eq:elbo}. The update formulas for $ \mathbf{W} $ and $ \mathbf{H} $ are then obtained by using standard multiplicative rules~\cite{FevotBD09}:
\begin{equation}
\mathbf{H} \leftarrow \mathbf{H} \odot  \frac{\mathbf{W}^\top \left( \Vb \odot
	\left(\mathbf{W} \mathbf{H}\right)^{\odot-2} \right)}{\mathbf{W}^\top \left(\mathbf{W} \mathbf{H}\right)^{\odot-1} },
\label{updateH}
\end{equation}
\begin{equation}
\mathbf{W} \leftarrow \mathbf{W} \odot \frac{ \left( \Vb \odot
	\left(\mathbf{W} \mathbf{H}\right)^{\odot-2} \right)\mathbf{H}^\top}{\left(\mathbf{W} \mathbf{H}\right)^{\odot-1}\mathbf{H}^\top  },
\label{updateW}
\end{equation}
where $ \Vb = \left[\sum_{m_t} r^m(m_t)(|x_{ft}-\eta_{ft}^{m_t}|^2+\nu_{ft}^{m_t})\right]_{(f,t)} $, and $\odot$ signifies entry-wise operation. 
The parameters of the HMM, i.e. $\lambda$ and $\tau$, are updated by the standard formulae using the joint posterior probabilities computed by the forward-backward algorithm in the E-m step.
The complete inference and enhancement algorithm is summarized in Algorithm~\ref{alg:prop}.
		
		
		

		
		
		

\begin{algorithm}[t]
	\caption{SwVAE}\label{alg:prop}
	\begin{algorithmic}[1]
		\State \textbf{Input:} Trained A-VAE and AV-VAE models, noisy STFT frames $\lk \vect{x}_t \rk_{t=1}^T$, and visual embeddings $\lk \vect{v}_t \rk_{t=1}^T$.
		\State \textbf{Initialize:}\vspace{-1mm}
		\begin{itemize}
		[leftmargin=*]
		\compresslist
			\item The latent codes $\{\zb_{{m}_t}^{(d)}\}_{d=1}^D$ via the VAE encoders.
			\item The parameters of $r^s(\vect{s}|\vect{m})$ using \eqref{eq:rs-params}.
			\item The posterior $r^m(\vect{m})$ uniformly.
			\item The parameters $\vect{W}$, $\vect{H}$, $\tau$ and $\lambda$ (randomly).
		\end{itemize} 
		\State \textbf{While} stop criterion not met \textbf{do}:\vspace{-1mm}
		\begin{itemize}
		[leftmargin=*]
		\compresslist
			\item \textbf{E-$\vect{z}$ step:} Using
			\eqref{eq:rz-update}.
			\item \textbf{E-$\vect{s}$ step:} Using \eqref{eq:rs-params}.
			\item \textbf{E-$\vect{m}$ step:} Compute $ {q}_{mt}=\frac{\exp(-g_t({m}_{t}))}{\sum_{m_t} \exp(-g_t({m}_{t}))}$ using \eqref{eq:gm}, and run the forward backward algorithm \cite{bishop06} to obtain the posterior probability $r^m(m_t)$ and the joint posterior probability $\zeta^m(m_{t-1},m_t)$.
			
		\item \textbf{M step:} Update $ \Wb, \Hb $ using \eqref{updateW} and \eqref{updateH}, and $\lambda,\tau$ using the standard formulae with $r^m$ and $\zeta^m$ \cite{bishop06}.
		\end{itemize}
        \State \textbf{End while}
		\State \textbf{Speech enhancement:} Using \eqref{eq:se}.
		
	\end{algorithmic}
\end{algorithm}

\begin{table*}[t]
\centering
	\caption{Average PESQ, SDR and STOI values of the enhanced speech signals. Here, ``clean'' and ``noisy'' refer to visual data.}\vspace{-2mm}
\resizebox{\textwidth}{!}{
\begin{tabular}{|l|c|c|c|c|c||c|c|c|c|c||c|c|c|c|c|}
\hline
 Measure & \multicolumn{5}{c||}{PESQ} & \multicolumn{5}{c||}{SDR (dB)} & \multicolumn{5}{c|}{STOI} \\
\hline
{SNR (dB)} & {-5} & {0} & {5} & {10} & {15} & {-5} & {0} & {5} & {10} & {15} & {-5} & {0} & {5} & {10} & {15} \\ \hline\hline
Input & 1.44 & 1.67 & 2.04 & 2.30 & 2.72 & -12.30 & -7.30 & -3.45 & 1.88  & 6.73    & 0.22        & 0.32        & 0.45       & 0.56       & 0.68                    \\ \hline\hline
\cite{SadeA19a} - clean & \textbf{1.70} & 1.92 & 2.29 & 2.48 & 2.66 & \textbf{-3.51} & 1.67 & 5.38 & 9.22 & 12.07 & 0.24         & 0.35        & 0.47       & 0.55       & 0.65 \\ \hline
SwVAE - clean  &  1.67 & \textbf{1.97} & \textbf{2.39} & \textbf{2.62} & \textbf{2.83} & -3.59 & \textbf{2.00} & \textbf{6.24} & \textbf{10.73} & \textbf{14.12} & {\textbf{0.25}} & {\textbf{0.36}} & {\textbf{0.51}} & {\textbf{0.61}} & {\textbf{0.72}}\\ \hline\hline
\cite{SadeA19a} - noisy & \textbf{1.66} & 1.91 & 2.22 & 2.41 & 2.51  & \textbf{-3.78} & 1.50 & 5.18 & 8.72 & 10.88 & 0.23                     & 0.34                   & 0.45                     & 0.53                     & 0.63 \\ \hline
SwVAE - noisy & 1.65 & \textbf{1.94} & \textbf{2.36} & \textbf{2.60} & \textbf{2.81} & -3.97 & \textbf{1.84} & \textbf{6.14} & \textbf{10.51} & \textbf{14.06} & {\textbf{0.24}} & {\textbf{0.35}}  & {\textbf{0.50}} & {\textbf{0.59}}  & {\textbf{0.67}} \\ \hline
\end{tabular}}
\label{tab:measures}\vspace{-3mm}
\end{table*}
\subsection{Novelty of SwVAE w.r.t.\ \cite{SadeA19a}}\label{sec:related}
The closest work to ours is \cite{SadeA19a}, which uses a mixture model, comprising an A-VAE and an AV-VAE, as the generative model of clean speech. Though sharing some similarities, there are several crucial differences between the two methods. First, here we assume a Markovian dependency on the switching variable that ensures smoothness over time. Second, in \cite{SadeA19a} the following variational factorization is proposed: $p(\vect{s},\vect{z},\vect{m}|\vect{x}) \approx r^s(\vect{s})r^z(\vect{z})r^m(\vect{m})$, where $r^s$ and $r^z$ are not conditioned on $\vect{m}$. This is in contrast to our proposed factorization given in \eqref{eq:var-dists}, which provides a more effective approximation and a robust initialization for the latent codes $\vect{z}$, as required by the inference algorithm. More precisely, in the proposed framework, the parameters of $r^s(\vect{s}|\vect{m})$ are initialized using its respective set of latent codes $\vect{z}$, which themselves are initialized by the corresponding encoders (see Section~\ref{sec:exp}), as opposed to \cite{SadeA19a} where a weighted combination of the latent codes (coming from different models) is used for initializing the parameters of $r^s(\vect{s})$. This might not be effective given that latent initialization is important in VAE-based AVSE \cite{SadeA20MinVAE}. Finally, the proposed posterior approximation $r^z(\vect{z}_t|{m}_t) = {\cal N}(\vect{c}_{tm},\vect{\Omega}_{tm})$ makes sampling, needed by \eqref{eq:gamma_}, more efficient than the method of \cite{SadeA19a} which relies on the computationally demanding Metropolis-Hastings algorithm~\cite{bishop06}.

\section{Experiments}\label{sec:exp}
\paragraph*{Protocol} We evaluate the performance of SwVAE and compare it with~\cite{SadeA19a} {using the same experimental protocol}. We used two VAE models (A-VAE and AV-VAE)\footnote{For A-VAE, the prior of $ \vect{z}_t $ is a standard normal distribution, and $\vect{\Sigma}_{m_t}$ is a function of only $\vect{z}_t$; see \eqref{eq:gen-model}.} from~\cite{sadeghiLAGH19}, trained on the NTCD-TIMIT dataset~\cite{Abde17}. The test set includes 9 speakers, along with their corresponding lip region of interest, with different noise types: \textit{\ac{LR}}, \textit{White}, \textit{Cafe}, \textit{Car}, \textit{Babble}, and \textit{Street}, and noise levels: $ \lk -5,0,5,10,15 \rk $~dB. From each speaker, we randomly selected 150 examples per noise level for evaluation.

The parameters for the algorithm of \cite{SadeA19a} where set as their proposed values. Both of the algorithms were run for $ 200 $ iterations, on the same test set. For optimizing \eqref{eq:rz-update}, the Adam optimizer \cite{kingma2014adam} was used with a learning rate of $0.05$ for 10 iterations. Moreover, we used $D=20$ samples to compute \eqref{eq:gamma_} and \eqref{eq:gm}. The $\vect{c}_{tm},\vect{\Omega}_{tm}$ parameters of $r^z$ were, respectively, initialized with the means and variances at the output of the respective VAE encoders by giving $(\vect{x}_t, \vect{v}_t)$ as their inputs. The parameters of $r^s$ are then initialized using \eqref{eq:rs-params} and \eqref{eq:gamma_}.

The two AVSE algorithms were run on the test set with both clean 
visual data as well as artificially generated noisy versions, where about one third of the total video frames per test instance were occluded. Similarly to \cite{SadeA19a}, the occlusions were simulated by random patches of standard Gaussian noise added to randomly selected sub-sequences of 20 consecutive video frames. We used three standard speech enhancement scores, i.e., \ac{SDR} \cite{vincent2006performance}, \ac{PESQ}  \cite{rix2001perceptual}, and \ac{STOI}~\cite{taal2011algorithm}. \ac{SDR} is measured in decibels (dB), and \ac{PESQ} and \ac{STOI} values lie in the intervals $[-0.5,4.5]$ and $ [0,1] $, respectively (the higher the better). 

\paragraph*{Results} Table~\ref{tab:measures} summarizes the results, averaged over all the test samples, for the three performance measures, and clean as well as noisy visual data. From this table, we can see that in terms of PESQ and SDR, SwVAE outperforms \cite{SadeA19a}, with the performance difference being more significant in high SNR values. In terms of the intelligibility measure, i.e., STOI, the proposed method exhibits much better performance than~\cite{SadeA19a}. These observations are consistent for both clean and noisy visual data. Furthermore, the two algorithms show robustness to noisy visual data, which is especially noticeable in terms of STOI. However, for the algorithm of~\cite{SadeA19a} the performance drop due to noisy visual data is higher than SwVAE. Supplementary materials are available online\footnote{\url{https://team.inria.fr/perception/research/swvae/}}.

\section{Conclusion}
In this paper, we proposed a noise-agnostic audio-visual speech generative model based on a sequential mixture of trained A-VAE and AV-VAE models, combined with an NMF model for the noise variance. The switching variable allows us to seamlessly use either of the auto-encoders for speech enhancement, without requiring supervision.
We detailed a variational expectation-maximization approach to estimate the parameters of the model as well as to enhance the noisy speech. 
The proposed algorithm, called switching VAE (SwVAE), exhibits promising performance when compared to the previous work \cite{SadeA19a} on robust AVSE. In the future, we would like to explore the use of Dynamical VAEs~\cite{girin2020dynamical} for unsupervised AVSE.

\balance
\bibliographystyle{IEEEbib} 
\bibliography{myref_compressed}

\end{document}

%% file: Header.tex
\newcommand{\lk}{ \left\{ }
\newcommand{\rk}{ \right\} }

\newcommand{\Rbb}{{\mathbb{R}}}

\newcommand{\Hb}{{\bf H}}

\newcommand{\zb}{{\bf z}}

\newcommand{\Wb}{{\bf W}}

\newcommand{\vb}{{\mathbf v}}
\newcommand{\Vb}{{\mathbf V}}

\newsavebox\mybox

